\documentclass[aps,twocolumn,prl,showpacs,footinbib,groupedaddress,superscriptaddress]{revtex4}
\usepackage{float}
\usepackage{amsfonts,amsmath,amssymb,times,bbm}
\usepackage{graphicx,subfigure}

\newcommand{\ket}[1]{|#1\rangle}
\newcommand{\bra}[1]{\langle#1|}

\newcommand{\eq}[1]{Eq.~(\ref{#1})}

\begin{document}

\title{Quantum Teamwork for Unconditional Multiparty Communication with Gaussian States}
\author{Jing Zhang}\email[E-mail: jzhang74@yahoo.com; jzhang74@sxu.edu.cn.]{}
\affiliation{State Key Laboratory of Quantum Optics and Quantum
Optics Devices, Institute of Opto-Electronics, Shanxi University,
Taiyuan 030006, P.R.China.}
\author{Gerardo Adesso}
\affiliation{School of Mathematical Sciences, University of Nottingham,
University Park,  Nottingham NG7 2RD, UK.}
\author{Changde Xie}
\affiliation{State Key Laboratory of Quantum Optics and Quantum
Optics Devices, Institute of Opto-Electronics, Shanxi University,
Taiyuan 030006, P.R.China.}
\author{Kunchi Peng}
\affiliation{State Key Laboratory of Quantum Optics and Quantum
Optics Devices, Institute of Opto-Electronics, Shanxi University,
Taiyuan 030006, P.R.China.}

\begin{abstract}

We demonstrate the capability of continuous variable Gaussian states to communicate multipartite quantum information. A quantum teamwork protocol is presented according to which an arbitrary possibly entangled multimode state can be faithfully teleported between two teams each comprising many cooperative users.
We prove that $N$-mode Gaussian weighted graph states exist for arbitrary $N$, that enable unconditional quantum teamwork implementations for any arrangement of the teams. These perfect continuous variable  maximally multipartite entangled resources are typical among pure Gaussian states and are unaffected by the entanglement frustration occurring in multiqubit states.

\end{abstract}


\maketitle

\noindent {\it Introduction.}---
Information technology is hectically evolving towards the quantum scale. The discovery that  entanglement, apart from being a fascinating founding feature of quantum theory, can be employed as a resource for novel or enhanced processing, manipulation, and distribution of information \cite{nielsenchuang} has spurred enormous theoretical and experimental progress.  An appealing perspective for future mainstream quantum information technology is to be implemented over a  ``quantum internet'' \cite{qinternet} architectured as a network where light buses and atomic storage units are interfaced to enable secure information transfer and quantum computation. The main routes towards the quantum internet are either relying on discrete-variable  interfaces between single photons and single trapped atoms, which are physical representations of qubits, or on continuous-variable (CV) interfaces between light modes and mesoscopic atomic ensembles, which are physical realizations of bosonic modes.
Historically, most of the concepts of quantum information and computation have
been first developed in the qubit regime, and more recently generalized to a CV scenario where very promising perspectives have been acknowledged concerning both
general theoretical insights and experimental realizations \cite{brareview},  due to the relative
simplicity and high efficiency in the generation, manipulation, and
detection of special classes of CV states, Gaussian states \cite{ourreview}.
However, the two scenarios for quantum technological implementations have mostly progressed in parallel, with no truly radical instance  in which the performance of one approach is demonstrated to be unique and  unmatched by the other.

Here we introduce a general multipartite communication primitive -- the `quantum teamwork' --  that cannot be implemented perfectly with $N$-qubit resources beyond the threshold $N\ge 8$. On the contrary, we identify families of  Gaussian multipartite CV entangled states of an arbitrary number of modes which remarkably enable the unconditional transmission of multimode quantum states between two teams of arbitrarily arranged parties over quantum networks. These CV resources, belonging to the class of Gaussian weighted graph states \cite{one,thirteen,twenty-one}, are shown to be typical in the space of pure Gaussian states, and are ideally unaffected by the kind of entanglement frustration that arises in complex low-dimensional many-body systems \cite{facchi}. We present examples and applications and we discuss the efficiency of quantum teamwork implementations in presence of realistic imperfections.
Our results demonstrate that at a fundamental level CV entanglement offers enhanced perspectives for  quantum technology.

\noindent {\it Quantum teamwork.}---
A quantum police department is investigating malicious activities over the quantum internet \cite{qinternet}. A briefing is issued to $N$ agents in the form of an entangled $N$-party quantum state $\varrho^{(0)}_N$ distributed by the central station,
possibly
supplemented with classically broadcast information.
The chief publicly orders the agents to split in two teams of $K$ and $N-K$ members, respectively, where we shall assume $K\le N-K$ here and in the following. The parties carry their share of the entangled resource and the two teams independently proceed in the investigation. They gather progress encoded in generally entangled states $\ket{\Phi}$ of up to $K$ subsystems, supplied by independent informants. Suppose team $A$ wishes to communicate some undisclosed information to team $B$. To do so, the members of each team must act cooperatively on their respective blocks of the original entangled system, in such a way that the distributed resource $\varrho_N^{(0)}$  is  converted into a perfect $K$-user quantum teleportation channel, i.e. a tensor product of $K$ maximally entangled states of two subsystems, each belonging to one of the teams. One team can then transfer with perfect fidelity the $K$-party state $\ket{\Phi}$, containing the investigation progress, to the other team which may process or store it for further inspection. The teams then meet up for a new briefing, another entangled resource $\varrho^{(1)}_N$ is redistributed and possibly a different configuration of the team squads is ordered for the subsequent stage of the investigation. And the process continues for $n$ steps until the case is closed.

We have just introduced an instance of {\it quantum ``teamwork''}. It is a very powerful and practical primitive that generalizes conventional quantum teleportation \cite{bennett,brakim} and multipartite teleportation networks  \cite{network} and promises to be of many uses in quantum technology implementations. Crucial for the feasibility of quantum teamworks are the multipartite entangled resources $\varrho^{(j)}_N$  distributed to the parties at each briefing. Above, we assumed different resources  be needed at each step $j$, adapted on the team configuration. This would however require some control at the station level, i.e. over the initialization of the protocol, posing a drawback to its versatility. We raise then an intriguing question: Do universal resources exist that enable the unconditional realization of quantum teamworks for {\em any} arrangement of the two teams? In mathematical terms, do $N$-partite quantum states $\varrho_N$ exist which are locally equivalent, with respect to any  $K|(N-K)$ bipartition,  to the tensor product of exactly $K$ maximally entangled states, plus $N-2K$ uncorrelated single-party states?

\noindent {\it Maximally multipartite entangled states.}---
Addressing such a question is quite challenging, mainly because the structural aspects of multipartite entanglement are still largely unknown \cite{rmps}. If one restricts the search for $\varrho_N$  to states of $N$ qubits, however, one has in general a negative answer.
Consider e.g. $N=4$:  while protocols for teleporting an arbitrary state of two qubits via either two Bell pairs \cite{rigolin} or a genuinely four-qubit multipartite entangled state \cite{yeo} exist, they cannot work for unit fidelity for all of the three possible splittings of four parties into two two-party teams.
The most suitable qubit resource states to implement quantum teamworks would be the  perfect ``maximally multipartite entangled states'' (MMES) recently studied by Facchi {\it et al.} \cite{facchi}, which are those $N$-qubit states states $\ket{\Psi}$ that can be written in Schmidt form as $\ket{\Psi}_{AB} = \sum_{j=1}^K \sqrt{1/K} \ket{\phi_j}_A \ket{\chi_j}_B$ across any $A|B$ partition where the block $A$ is made of $K$ qubits.  They have the defining property of exhibiting the maximum allowed bipartite entanglement for all possible splittings.   However, it has been proven that qubit perfect MMES do not exist for $N=4$ and for all $N\ge8$, indicating the strong presence of an entanglement frustration as complexity increases \cite{facchi}. Nontrivial quantum teamwork instances can hence be faithfully implemented only among $5$ or $6$ parties (the case $N=7$ is still unsettled \cite{facchi}) when each party is allotted a single qubit.

\noindent {\it Perfect maximally multipartite entangled states for continuous variables.}---
In a CV scenario, one could translitterate the notion of MMES in a weak sense, by looking for states whose bipartite entanglements are simultaneously diverging across all bipartitions. It is easy to find CV states which meet these requirements, the most prominent examples being  the GHZ-type (GHZ stands for Greenberger-Horne-Zeilinger) permutation-invariant  Gaussian states \cite{network} of $N$ modes, in the limit of infinite local squeezing. However, for any bipartition these states are locally equivalent to a single maximally entangled state [Einstein-Podolsky-Rosen (EPR) pair] and as such only enable a very limiting case of quantum teamwork, where $K=1$. This special instance of teleporting arbitrary single-mode states from any sender to any receiver, the latter acting cooperatively with $N-2$ assisting parties, was already introduced as a `quantum teleportation network' \cite{network} and has been demonstrated experimentally for $N=3$ \cite{naturusawa}.

In this Letter we go beyond these limitations by identifying $N$-mode CV Gaussian resources which are locally equivalent to a maximal number of EPR pairs for any partition, thus enabling unconditional quantum teamworks, and properly embodying the stronger notion of  `perfect' CV MMES.  Precisely,  we define a perfect CV MMES  as a (not necessarily Gaussian) $N$-mode state such that, with respect to any $A|B$ partition where the block $A$ is made of $K$ modes, it can be transformed by local unitaries into the tensor product of exactly $K$ EPR pairs (and $N-2K$ uncorrelated single-mode states). Clearly such a state has infinite energy and stands as an idealized unphysical limit, therefore the notion of perfect CV MMES has to be understood in the sense of a family of states,  depending on a parameter (for instance, a local squeezing degree), which converge asymptotically  to the ideal perfect CV MMES when the parameter is very large. Each member of the family, characterized by a finite value of the parameter, would be an imperfect approximation of the perfect CV MMES, but crucially this approximation can be made arbitrarily precise.

We shall look for perfect CV MMES within the family of Gaussian weighted graph states of a $N$-mode CV system. Gaussian states, up to local displacements, are completely specified by the covariance matrix $\Gamma$ of the second canonical moments \cite{ourreview}.
A weighted graph Gaussian state $\ket{\Psi}$ is a  pure $N$-mode Gaussian state whose covariance matrix is defined in terms of a local squeezing degree $r>0$, and associated to a weighted $N$-vertex graph ${\cal G}$ with adjacency matrix $\Omega$ via a canonical prescription \cite{one,thirteen,twenty-one}, as detailed in the Supplementary material \cite{noteappendix}.
For any splitting of the modes into a $K$-mode block owned by party $A$ and a $(N-K)$-mode block owned by party $B$ (recall that $K \le N-K$), any pure $N$-mode Gaussian state $\ket{\Psi}_{AB}$ is equivalent, up to local (with respect to the bipartition) Gaussian unitary operations, to the tensor product of $M \le K$ entangled two-mode squeezed states $\ket{\psi(r_{i})}_{i\in A,i+K \in B}$ and additional $N-2M$ uncorrelated single-mode vacua \cite{boteroecc}. The two-mode squeezed states $\ket{\psi(r)} = \sum_n
\lambda^n \sqrt{1-\lambda^2} \ket{n,n}$ (with $\lambda=\tanh r$) are  prototypical CV   states which reproduce infinitely entangled EPR pairs in the limit $r \rightarrow \infty$. The number $M$ of entangled two-mode squeezed states in the normal form of a pure $N$-mode Gaussian state for a given partition is given by the symplectic rank  of its reduced $K$-mode covariance matrix \cite{noterank}.

In the case of weighted graph Gaussian states,
by writing the adjacency matrix $\Omega$ in block form with respect to the $A|B$ partition, $\Omega = {{\Omega_A \ \ \  \Omega_{AB}}\choose{\Omega_{AB}^T \ \Omega_B}}$, one has $M(\ket{\Psi}_{AB})={\rm rank}(\Omega_{AB})$ \cite{one,twenty-six}. Moreover, by construction, for these states, the two-mode squeezed pairs appearing in the normal form all simultaneously converge to EPR pairs for a diverging $r$. Thus we obtain the following powerful and simple criterion: a pure $N$-mode Gaussian weighted graph state with adjacency matrix $\Omega$ is a perfect CV MMES in the limit of infinite local squeezing if and only if all the $K \times (N-K)$ off-diagonal block submatrices $\Omega_{AB}$ have matrix rank $K$.

\begin{figure}[t]
 \subfigure[] {\includegraphics[width=2.3cm]{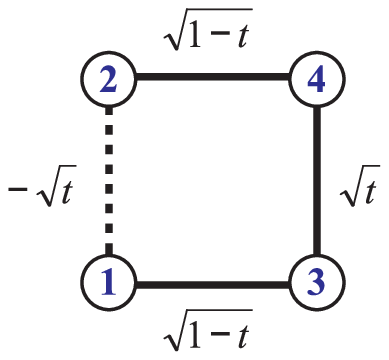}} \hspace{.2cm}
\subfigure[] {\includegraphics[width=6cm]{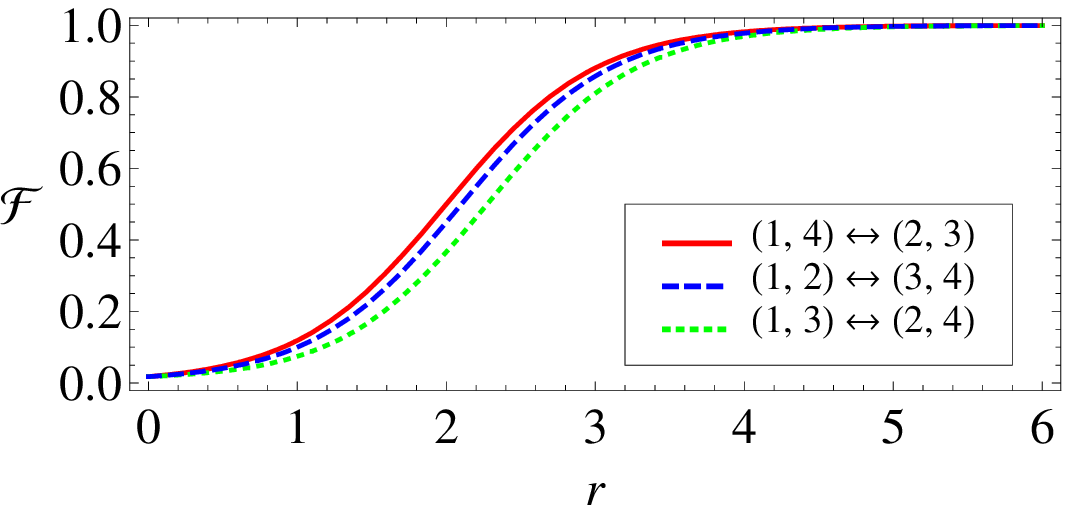}}

\caption{(Color online). (a) Weighted graph definition of four-mode perfect CV MMES locally equivalent to the state ${\ket{\Psi^{(4)}}}$;  the weights displayed on the graph are the elements of the adjacency matrix. (b) Fidelity of the quantum teamwork for teleporting input two-mode squeezed states with squeezing $z=2$, plotted versus the resource squeezing $r$ for the three possible arrangements of sender and receiver teams (the beam splitter transmittivity is $t=1/3$). \label{figfid} }
\end{figure}

\noindent {\it How quantum teamwork works.}---
We are ready to provide examples of Gaussian resources useful for quantum teamworks. We first observe that the direct analogues of perfect qubit MMES for $N=5,\,6$ \cite{facchi}, described by  unweighted, partially connected  graphs,
are perfect CV MMES, as one would expect.
For $N=4$ no perfect qubit MMES exist \cite{facchi}. However, we disclosed several instances of four-mode CV perfect MMES. One such instance is provided by (the infinite-squeezing limit of) a  Gaussian state $\ket{\Psi^{(4)}}$ with irreducible three-party correlations which has been recently introduced  \cite{twenty-nine,thirty} and experimentally demonstrated \cite{thirty-one}. It is defined by the action of a beam splitter on two beams each taken from a two-mode squeezed state, $\ket{\Psi^{(4)}}=U(t)_{23}[\ket{\psi(r)}_{12} \otimes \ket{\psi(r)}_{34}]$, where $U(t)$ is a beam splitter transformation with transmittivity $t$. This state is locally equivalent to the four-mode ring weighted CV graph state depicted in Fig.~\ref{figfid}(a).

For the  resource $\ket{\Psi^{(4)}}$ it is instructive to illustrate the explicit steps needed for the teleportation of arbitrary two-mode states from any two parties paired as team $A$ to the remaining two parties of team $B$. If $A=(1,4)$ the only local operation required to obtain 2  two-mode squeezed entangled pairs  is obviously an inverse beam splitter with transmittivity $t$ on the $B$ modes. If $A=(1,2)$ both teams have to locally mix their respective modes at a $50:50$ beam splitter, and further apply local squeezing operations,  with squeezing $s$ on modes $1$ and $4$, and $1/s$ on modes $2$ and $3$ ($s=[(\cosh 2r - \sqrt t \sinh 2r)/(\cosh 2r - \sqrt t  \sinh 2r)]^{1/4}$), to reproduce a dual entangled channel. If $A=(1,3)$ the same steps are required but with the following modifications, $t \rightarrow 1-t$ and $2 \leftrightarrow 4$. Suppose now a random arrangement is chosen for the two teams, and team $A$ wishes to teleport an arbitrary  two-mode (entangled) state, e.g. a two-mode squeezed state $\ket{\psi(z)}_{in}$, to team $B$. After the local operations described above, each input mode is independently teleported according to the conventional Braunstein-Kimble scheme \cite{brakim,twenty-seven} using the corresponding two-mode squeezed entangled pair \cite{adhikariecc}, and finally team $B$ obtains the two-mode state $\varrho_{out}$. The fidelity ${\cal F}={_{in}\bra{\psi(z)}} \varrho_{out} \ket{\psi(z)}_{in}$ between input and output reads
\begin{equation}\label{fidelity4}
{\cal F}=  \frac{\exp(2 r_A)}{2(\cosh 2r_A+\cosh 2z)}\,,
\end{equation}
depending on  both the input squeezing $z$ and the effective  squeezing $r_A$ of the two entangled pairs in the normal form of $\ket{\Psi^{(4)}}$ relative to the partition $A|B$. We have  $r_{(1,4)}=r$, $r_{(1,2)} = \frac12 {\rm arccosh}\sqrt{\cosh^2 2r-t \sinh^2 2r}$, $r_{(1,3)} = \frac12 {\rm arccosh}\sqrt{\cosh^2 2r-(1-t) \sinh^2 2r}$. For any finite $z$, $0<t<1$, and $r \rightarrow \infty$, in which limit the entangled channels reproduce exactly EPR pairs, the fidelity converges to unity for all bipartitions [see Fig.~\ref{figfid}(b)].  This is an {\em unconditional} realization of quantum teamwork communication for $N=4$, experimentally feasible with current optical technology \cite{twenty-seven,thirty-one,naturusawa}.

More examples of perfect CV MMES are provided in the Supplementary material \cite{noteappendix}.

\noindent {\it Typicality of continuous variable perfect maximally multipartite entangled states.}---
We may now attempt an estimate of the volume occupied by perfect CV MMES in the space of  pure Gaussian states for arbitrary $N$. The reduced symplectic eigenvalues $\nu_j^{(A)} \ge 1$ ($j=1,\ldots,K$) of the $K$-mode covariance matrix $\Gamma_A^{(N)}$ of a generic $N$-mode pure-state covariance matrix $\Gamma_N$, corresponding to any block $A$,  are a continuous analytic function of the real  elements of the covariance matrix \cite{ourreview}. It follows that the set of states which certainly fail to be  perfect MMES, characterized by  one or more $\nu_j^{(A)}=1$, is of null measure. If we specialize to weighted graph Gaussian states, where by construction if all $\nu_j^{(A)} > 1$ then they all diverge for infinite local  squeezing, we can conclude that perfect CV MMES families are henceforth {\it typical} \cite{serafozzi} in the space of (weighted graph) pure Gaussian states, in the sense that randomly picked (weighted graph) pure Gaussian states for an arbitrary number of modes are expected to reproduce perfect MMES for diverging local squeezing \cite{notequbit}.
We tested the typicality argument by a numerical construction of pure $N$-mode Gaussian states described by (generally complete) graphs associated to a symmetric adjacency matrix $\Omega$ with null principal diagonal and  random off-diagonal integer weights $\pm \Omega_{ab} \in [0, N]$. Remarkably,  we  found instances of perfect CV MMES at the  first run for each $N$ (we tested up to $N=100$). An example for $N=20$ is presented in the Supplementary material \cite{noteappendix}.
As  mentioned, notable exceptions lying in the null-measure set of non-perfect MMES are the CV GHZ-type states \cite{network},  associated to fully connected unweighted graphs.  As soon as the  symmetry is broken and some randomness is plugged in the interaction weights, perfect maximal multipartite entanglement pronouncedly arises in CV Gaussian states, which are thus unaffected by the entanglement frustration featuring in complex many-body low dimensional systems   \cite{facchi}.

\noindent {\it Applications and remarks on practical implementations.}---
Perfect CV MMES stand as promising resources for operating the  ``quantum internet'' \cite{qinternet}.  All Gaussian weighted graph states,  including perfect CV MMES, are efficiently engineerable with off-line squeezers and linear optics \cite{fourteen}.
Let us also recall that several classes of cluster states, belonging to the family of  (weighted) graph Gaussian states \cite{thirteen,twenty-one}, are proven  universal resources for CV  one-way quantum computation 
\cite{seventeen}. Given the typicality of perfect CV MMES, we may expect that suitable graph structures exist for arbitrarily large $N$  such  to achieve both the universality for quantum computation \cite{fundamentals} and the CV MMES property. By construction, perfect CV MMES enable the creation of  EPR pairs between any two modes, and have unbounded entanglement width scaling  linearly with $N$, thus fulfilling the  necessary conditions for efficient universality \cite{fundamentals}.
  Remarkably, perfect CV MMES automatically enable faithful ``disk operations'' such as cut \& paste of large sectors of stored data. Here the input data are
 encoded in the state of multiple atomic ensembles, team $A$ owns the light modes interacting with the source  units, and team $B$ controls the target storage units. Building on the demonstrations of quantum memory and  teleportation from light to matter and viceversa \cite{polzikhonda}, unconditional reallocation of multimode  data (up to half the total disk space) can be performed with our scheme.
Perfect CV MMES for $N$ modes enable moreover data compression in the form of  multipartite dense coding to transmit $N$-way classical signals simultaneously \cite{twenty-eight}. Further  applications include quantum secret/state sharing \cite{sharing}.

Clearly, any implementation will be affected by experimental imperfections, mainly traceable to the usage of a finite degree of squeezing in the preparation of the teamwork resources, which are  in practice imperfect approximations of perfect CV MMES. This results in a non-unit fidelity in the transmission (as seen in Fig.~\ref{fidelity4}(b) for finite $r$, where e.g ${\cal F} \approx 80\%$ for $20$~dB of resource squeezing). The ensuing inefficiency of the protocol propagates with the number $K$ of input modes to be teleported between the two teams. For instance, suppose the teamwork resource is a  weighted graph state of $N=2K$ modes, locally reducible to a tensor product of $K$ two-mode squeezed states, with squeezing $r$, across any partition into two blocks of $K$ modes each. Let us consider as input state a $K$-mode GHZ-type symmetric state with local squeezing $z$ \cite{network,adhikariecc}. Each mode of the input state is individually teleported via a corresponding approximate EPR channel. The overall teamwork fidelity ${\cal F}_K$ scales exponentally with $K$, as $[(\cosh z)^{-1}  {\cal F}_1]^K \le {\cal F}_K \le {{\cal F}_1}^K$, where ${\cal F}_1 = (1+\tanh r)/2$ is the  fidelity \cite{brakim} for teleporting a single-mode vacuum via a two-mode squeezed state with squeezing $r$. We observe that if $r \gg 0$ (quasi-perfect CV MMES resource) then the performance is slightly affected by  a larger input size, while for moderate (realistic) squeezing $r$ in the resource the performance can be quickly degraded with $K$, and the degradation is  enforced by the amount of entanglement in the input state (in this case, $z$). Suitable generalizations and optimizations of the conventional Braunstein-Kimble teleportation scheme \cite{brakim}, which is a building block of our networked scheme, might result in improved efficiencies for a given, finite resource squeezing degree. However, this is beyond the scope of this work, and we remark that the conceptual relevance of our results stays solid.

The demonstration advanced in this Letter of the   powerful features of quantum teamwork  protocols,  peculiar of CV  systems and  structurally precluded to  many-qubit scenarios, opens an avenue for the deep investigation of the characteristic traits of high-dimensional multipartite entanglement, and their exploitation to convey multimedia information over a global quantum communication  web.


 \noindent We acknowledge support  by NSFC for Distinguished Young
Scholars (Grant No.\ 10725416), National Basic Research Program of
China (Grant No.\ 2006CB921101), NSFC Project for Excellent Research
Team (Grant No.\ 60821004), NSFC (Grant No.\ 60678029, 60736040). We thank A. Serafini and N. Menicucci for fruitful discussions.

\clearpage
\appendix
\section*{Supplementary material}

\subsection{S1. Weighted graph Gaussian states}

A weighted graph quantum
 state is described by a mathematical graph ${\cal G}=({\cal V},{\cal E})$, i.e. a
 set of $N$ vertices ${\cal V}$ connected by a set of edges ${\cal E}$
\cite{one}, in which every edge is tagged by a factor
$\Omega_{ab}$ reflecting the interaction strength between
vertices $a$ and $b$.
For CV systems, the Weyl-Heisenberg group of phase-space displacements is a Lie group with generators
$\hat{x}=(\hat{a}+\hat{a}^\dagger)/\sqrt{2}$ (position) and $\hat{p}=-i(\hat{a}-\hat{a}^\dagger)/\sqrt{2}$
(momentum), where
$[\hat{x},\hat{p}]=i$ (with $\hbar=1$). The single-mode Pauli
operators (so termed in analogy with qubit systems) are defined as $X(s)=\exp[-is\hat{p}]$ and
$Z(t)=\exp[it\hat{x}]$ with $s,t\in \mathbb{R}$.  In the preparation procedure of CV weighted graph
states \cite{twenty-one}, only local Clifford (Gaussian) operations are used:
first, one prepares each mode (or graph vertex) in a momentum-squeezed
state (analogue of Pauli-$X$
eigenstate), then, one applies a quantum non demolition (QND) interaction
($C_{Z}(\Omega)=\exp[i\Omega\hat{x}_{1}\bigotimes\hat{x}_{2}]$) with
 coupling  strength $\Omega_{jk}$ to each connected pair of
modes $(j,k)$.
The resulting CV Gaussian weighted graph state becomes, in the
limit of infinite local squeezing, a zero-eigenvalue simultaneous eigenstate of the position/momentum linear combination operators $\hat{g}_{a}=(\hat{p}_{a}-\sum_{b\in
N_{a}}\Omega_{ab}\hat{x}_{b})$, where  $a\in {\cal V}$
and modes
 $b\in {\cal N}_{a}$ are the nearest neighbors (connected vertices) of mode $a$.

In general, the canonical definition of a weighted graph pure Gaussian state $\ket{\Psi}$ in terms of the adjacency matrix $\Omega$ is the following. Letting $r>0$ be  a local squeezing degree, we define  $\ket{\Psi}$  as the Gaussian state whose field operators have the following variances: ${\rm Var}(\hat{p}_{a}-\sum_{b\in
{\cal N}_{a}}\Omega_{ab}\hat{x}_{b})=e^{-2r}$, ${\rm Var}(\hat{x}_a) = e^{2r}$, $\forall a=1,\ldots,N$. The Wigner distribution of $\ket{\Psi}$ can be constructed as $W(R)=\pi^{-N} \exp[-R^T \cdot \Gamma^{-1} \cdot R] = \pi^{-N} \exp\{-e^{2r}[\sum_a({p}_{a}-\sum_{b\in
{\cal N}_{a}}\Omega_{ab}{x}_{b})^2]-e^{-2r}[\sum_a x_a^2]\}$, where $R=(x_1,\,p_1,\,x_2,\,p_2,\,\ldots,\,x_N,\,p_N)^T$ is the vector of phase-space variables. This implicitly defines the covariance matrix $\Gamma$.

\subsection{S2. Further examples of perfect CV MMES}

Here we wish to provide a general analytical family of CV perfect MMES $\ket{\Psi^{(N)}}$ for  $N \le 12$ modes, which are inequivalent to the examples presented in the main text. These states are defined in the graph formalism by an adjacency matrix which is of Toeplitz form
\begin{equation}\label{toeplitz}\Omega^{(N)}=
{{
\mbox{$\left(
\begin{array}{lllll}
 a_1 & a_2 & \cdots  & a_{N-1} & a_N \\
 a_2 & a_1 & a_2 & \ddots & a_{N-1} \\
 \vdots  & a_2 & \ddots & \ddots & \vdots  \\
 a_{N-1} & \vdots  & \ddots & a_1 & a_2 \\
 a_N & a_{N-1} & \cdots  & a_2 & a_1
\end{array}
\right)$}}}\,,
\end{equation}
where $a_{j^{\rm odd}}=0$ and $a_{j^{\rm even}}=(-1)^{j/2+1} (j^2/4)$.
It can be readily checked that no  frustration is present in these Gaussian states (see Supplementary Section S3), which enable -- for a diverging local squeezing -- the unconditional quantum teamwork communication of up to $6$-mode states  with perfect fidelity between two teams of arbitrarily arranged parties.

We also wish to present an explicit example of perfect CV MMES of a higher number of modes (specifically, $N=20$) as obtained from a random construction of the adjacency matrix with the aim of testing the typicality of perfect CV MMES (as detailed in the main text).
A single iteration of our numerical program has provided the following adjacency matrix $\Omega$

\begin{widetext}
\begin{equation} \label{azz}
\Omega={\footnotesize{\begin{array}{c}\left(
\begin{array}{cccccccccccccccccccc}
 0 & 1 & -1 & -3 & 16 & -19 & -16 & -5 & 7 & 13 & -5 & -5 & -11 & -5 & 4 & -5 & 17 & 15 & 20 & -10 \\
 1 & 0 & -17 & -3 & 13 & -19 & 20 & 2 & -20 & 20 & 5 & 10 & -5 & 3 & -1 & -7 & 14 & -13 & -3 & -18 \\
 -1 & -17 & 0 & 0 & 0 & 7 & 0 & -12 & -17 & -12 & -14 & 2 & 1 & 6 & -7 & 12 & -10 & 4 & -11 & 0 \\
 -3 & -3 & 0 & 0 & 6 & -14 & 9 & 19 & -13 & 2 & 19 & -4 & -6 & -14 & -13 & 7 & 4 & 17 & 19 & 6 \\
 16 & 13 & 0 & 6 & 0 & -11 & -20 & 8 & -7 & 8 & 11 & 20 & 8 & -13 & 2 & 19 & -15 & -11 & -14 & -9 \\
 -19 & -19 & 7 & -14 & -11 & 0 & 10 & -6 & 5 & -4 & 7 & 7 & -7 & -1 & 20 & 10 & 9 & -16 & -5 & 9 \\
 -16 & 20 & 0 & 9 & -20 & 10 & 0 & -15 & 8 & 1 & 19 & 0 & 10 & 8 & -13 & 1 & -13 & -2 & -7 & 10 \\
 -5 & 2 & -12 & 19 & 8 & -6 & -15 & 0 & 20 & 5 & -13 & 20 & 19 & -20 & -18 & -17 & 9 & -6 & -13 & 9 \\
 7 & -20 & -17 & -13 & -7 & 5 & 8 & 20 & 0 & 0 & 5 & -11 & 1 & 7 & -5 & -3 & -17 & -7 & 6 & 1 \\
 13 & 20 & -12 & 2 & 8 & -4 & 1 & 5 & 0 & 0 & 9 & -20 & -2 & -17 & -16 & -11 & -14 & -3 & 12 & 14 \\
 -5 & 5 & -14 & 19 & 11 & 7 & 19 & -13 & 5 & 9 & 0 & -10 & 7 & -20 & -12 & -12 & -18 & -1 & 14 & -1 \\
 -5 & 10 & 2 & -4 & 20 & 7 & 0 & 20 & -11 & -20 & -10 & 0 & 10 & -1 & 2 & -18 & -4 & -1 & 16 & 0 \\
 -11 & -5 & 1 & -6 & 8 & -7 & 10 & 19 & 1 & -2 & 7 & 10 & 0 & -7 & 16 & 8 & 1 & 18 & -5 & 3 \\
 -5 & 3 & 6 & -14 & -13 & -1 & 8 & -20 & 7 & -17 & -20 & -1 & -7 & 0 & -3 & 8 & 5 & -12 & -5 & 1 \\
 4 & -1 & -7 & -13 & 2 & 20 & -13 & -18 & -5 & -16 & -12 & 2 & 16 & -3 & 0 & 12 & -1 & -6 & 11 & 7 \\
 -5 & -7 & 12 & 7 & 19 & 10 & 1 & -17 & -3 & -11 & -12 & -18 & 8 & 8 & 12 & 0 & 20 & 9 & 8 & 9 \\
 17 & 14 & -10 & 4 & -15 & 9 & -13 & 9 & -17 & -14 & -18 & -4 & 1 & 5 & -1 & 20 & 0 & 7 & 0 & -12 \\
 15 & -13 & 4 & 17 & -11 & -16 & -2 & -6 & -7 & -3 & -1 & -1 & 18 & -12 & -6 & 9 & 7 & 0 & 9 & 6 \\
 20 & -3 & -11 & 19 & -14 & -5 & -7 & -13 & 6 & 12 & 14 & 16 & -5 & -5 & 11 & 8 & 0 & 9 & 0 & 2 \\
 -10 & -18 & 0 & 6 & -9 & 9 & 10 & 9 & 1 & 14 & -1 & 0 & 3 & 1 & 7 & 9 & -12 & 6 & 2 & 0
\end{array}
\right)\end{array}}}.
\end{equation}
\end{widetext}

\begin{figure}[t]
\includegraphics[width=8cm]{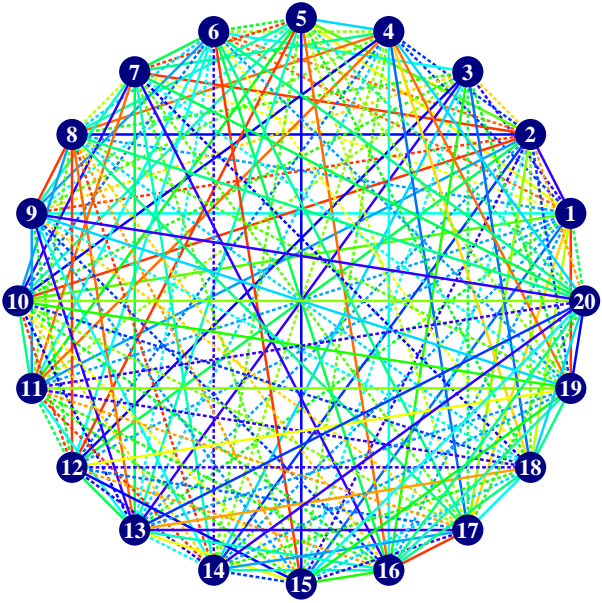}
\caption{Graph description of a  $20$-mode perfect CV MMES with randomly constructed adjacency matrix $\Omega$ given by \eq{azz}. Solid lines denote positive weights, while dotted lines are associated to negative weights. The absolute weights of the edges are integer numbers ranging from $1$ (blue) to $20$ (red).}\label{fig20}
\end{figure}

It can be checked (see Supplementary Section S3) that the associated Gaussian state is a perfect CV MMES in the limit of infinite local squeezing. The graph associated to the adjacency matrix of \eq{azz} is depicted in Fig.~\ref{fig20}

\subsection{S3. Verification of perfect CV MMES}

The perfect MMES property was checked through the reduced adjacency rank criterion.
At the  level of the covariance matrix, for any $A|B$  partition where $A$ is a block of $K \le \lfloor N/2 \rfloor$ modes, the considered states are locally equivalent to the tensor product of $N-2K$ uncorrelated vacua and  $K$ two-mode squeezed states  with squeezing degrees  $r_j^{(A)} = \frac12 {\rm arccosh}\sqrt{1+\alpha_j^{(A)}\ e^{4r}}$ where  $\alpha_j^{(A)}>0$ $\forall j=1,\ldots,K$.
 This means that in the limit $r \rightarrow \infty$ these resources have the maximal number of EPR pairs in their normal form with respect to all bipartitions.
 The local operations needed to transform the perfect CV MMES into a product of EPR pairs can be deduced case by case  from local graph complementation rules \cite{twenty-one}.


\begin{thebibliography}{99}
\bibitem{noterank} The symplectic rank of a covariance matrix  is the number of its symplectic eigenvalues strictly bigger than $1$ \cite{ourreview}.
\bibitem{notequbit}
In contrast, for $N$-qubit perfect MMES all the eigenvalues of any  $K$-qubit reduced density matrix must be exactly equal to $1/K$ \cite{facchi}, i.e.,  these states fill a null measure set (strictly empty for  $N=4$ and $N\ge 8$).
\end{thebibliography}
\end{document}